# Investigating the effect of non-resonant background variation on the CARS data analysis and classification

**RAJENDHAR JUNJURI,**[1,2] **TOBIAS MEYER-ZEDLER,**[1,2] **JÜRGEN POPP,**[1,2] **THOMAS BOCKLITZ**[1,2,*]

[1]*Leibniz Institute of Photonic Technology, Member of Leibniz Health Technologies, Member of the Leibniz Centre for Photonics in Infection Research (LPI), Albert-Einstein-Strasse 9, 07745 Jena, Germany.*
[2]*Institute of Physical Chemistry (IPC) and Abbe Center of Photonics (ACP), Friedrich Schiller University Jena, Member of the Leibniz Centre for Photonics in Infection Research (LPI), Helmholtzweg 4, 07743 Jena, Germany*
*\*Thomas.bocklitz@uni-jena.de*

**Abstract:** Non-resonant background (NRB) plays a significant role in coherent anti-Stokes Raman scattering (CARS) spectroscopic applications. All the recent works primarily focused on removing the NRB using different deep learning methods, and only one study explored the effect of NRB. Hence, in this work, we systematically investigated the impact of NRB variation on Raman signal retrieval. The NRB is simulated as a linear function with different strengths relative to the resonant Raman signal, and the variance also changed for each NRB strength. The resonant part of nonlinear susceptibility is extracted from real experimental Raman data; hence, the simulated CARS data better approximate the experimental CARS spectra. Then, the corresponding Raman signal is retrieved by four different methods: maximum entropy method (MEM), Kramers-Kronig (KK), convolutional neural network (CNN), and long short-term memory (LSTM) network. Pearson correlation measurements and principal component analysis combined with linear discriminant analysis (PCA-LDA) modelling revealed that MEM and KK methods have an edge over LSTM and CNN for higher NRB strengths. It is also demonstrated that normalizing the input data favors LSTM and CNN predictions. In contrast, background removal from the predictions significantly influenced Pearson correlation but not the classification accuracies for MEM and KK. This comprehensive study is done for the first time to the best of our knowledge and has the potential to impact the CARS spectroscopy and microscopy applications in different areas.

## 1. Introduction

Coherent Anti-Stokes Raman Scattering (CARS) is a third-order nonlinear spectroscopic technique that measures the vibrational response of the molecules[1, 2] and provides molecular fingerprint information [3, 4]. The ability to provide high-resolution, high-speed (~ ms)[5], and label-free imaging[6] are its advantages. In the CARS experiment, three laser beams (pump, stokes, and probe) are focused onto the target molecule, and a coherent anti-Stokes signal is generated with orders of magnitude enhancement in signal intensity compared to spontaneous Raman scattering[7]. The intensity of the CARS spectral line ($I_C$) can be described as

$$I_C(\nu) \, \alpha \, |\chi^{(3)}|^2 . I_S . I_p^2 . \tag{1}$$

Where, $\chi^{(3)}$ is the third order nonlinear susceptibility, $I_S$ and $I_p$ are the intensity of the Stokes and pump laser beams, respectively. Also, the nonlinear susceptibility $\chi^{(3)}$ is expressed as the sum of the non-resonant part $\chi^{(3)}_{NR}$ and the resonant part ($\chi^{(3)}_R$), and it can be written as

$$\chi^{(3)} = \chi^{(3)}_{NR} + \chi^{(3)}_R. \tag{2}$$

The $\chi^{(3)}_{NR}$ is purely a real term and appears due to the electronic contribution whereas $\chi^{(3)}_R$ is complex term corresponds to the resonant Raman signal.

The enhanced signal strength and other attractive features promoted CARS as a powerful spectroscopic tool for probing molecular structures with great precision. Moreover, the recent advancements in laser technology, optics, and data analysis have propelled CARS into the forefront of modern scientific investigations such as biological tissue mapping[8-10], lipid droplet dynamics[11], polymer blend analysis[12], etc[13]. Despite numeric advantages, the application of CARS remains non-trivial due to the inherent contribution of the non-resonant background (NRB) in the measured CARS signal, which significantly distorts the spectral line shapes. Several optical approaches have been demonstrated to reduce the effect of NRB but at the cost of increasing experimental complexity[14-17]. All these alternatives not only reduce the NRB but also the intensity of CARS signal. Hence, other numerical phase retrieval approaches, such as the maximum entropy method (MEM)[18], Kramers-Kronig (KK)[19], and wavelet prism analysis[20], are introduced to extract the resonant signal from the CARS spectra [21]. New correction approaches are also proposed to improve their performance[22-24]. However, these techniques require parameter optimization and reference NRB spectrum, which would become tedious for imaging applications.

A promising way to overcome the abovementioned complications is by utilizing deep learning (DL) algorithms, which have been recently explored in the CARS spectroscopy domain. SpecNet is the first work reported using DL models based on convolutional neural network (CNN) for NRB removal[25]. It has shown promising potential for removing NRB and paved the pathways for researchers. Besides, autoencoders[26], long short-term memory (LSTM) networks [27], and Bi-LSTM networks [28] have been explored by several research groups to tackle the issue of NRB. All these works have demonstrated that there is a growing interest in harnessing the capabilities of DL algorithms to overcome the hurdle of NRB.

Recently, Ryan et al. proposed that a physics-informed neural network improves their previous model efficiency based on autoencoders [29]. The authors have simulated the training dataset by considering the effect of laser pulse characteristics on the spectral shape, which more accurately mimics the spectrum close to the actual scenario. Earlier, we demonstrated that training the same SpecNet CNN model with semi-synthetic data[30] and fine-tuning/transfer learning [31] significantly improves its performance. These works show that DL model efficiency is greatly affected by the training dataset, which in turn depends on resonant signal and NRB, as mentioned in equations 1 and 2. In our recent work, we have also shown how different types of NRBs affect the DL model's performance[32].

In short, all recent studies have mainly focused on applying different DL algorithms for NRB removal. However, no study has investigated the influence of NRB variation on CARS data analysis to the best of our knowledge. Therefore, the present work aims to investigate the effect of NRB variation and the consequences on the corresponding Raman signal retrieval are presented in more detail. The NRB is assumed to as a linear function, and the CARS data is generated with different NRB strengths and variances. Here, the strength refers to how strong the NRB is relative to the resonant Raman signal. Also, the resonant part of nonlinear susceptibility is extracted from real experimental Raman data [30, 31]; hence, the simulated CARS data better approximate the experimental CARS spectra. Then, the corresponding Raman signal is extracted by four methods/models in which the first two are numerical phase retrieval approaches, MEM[18] and KK[19]. The remaining two are recently explored DL learning algorithms, SpecNet CNN[25] and LSTM[27]. Finally, we present a comparative study of the imaginary part prediction capabilities of these four models for different NRB strengths.

## 2. Materials and methods

This section presents a detailed overview of the CARS data simulation procedure. First, a brief overview of the experimental Raman data is presented in section 2.1. The NRB simulation details are discussed in section 2.2. Final CARS data simulation details are given in section 2.3.

*2.1 Experimental Raman data*

The Raman spectra of the six different bacteria samples were considered, and the details are given in Supplementary Table S1. The sample preparation procedure and experimental details are described here[33]. The six bacterial species are Escherichia coli DSM 423, Klebsiella terrigena DSM 2687, Pseudomonas stutzeri DSM 5190, Listeria innocua DSM 20649, Staphylococcus warneri DSM 20316, and Staphylococcus cohnii DSM 20261, from Deutsche Sammlung von Mikroorganismen and Zellkulturen GmbH (DSMZ). All these samples were cultivated in nine independent biological replicates. The suspension (1 μL) was placed on nickel foil and air-dried at room temperature prior to acquiring the Raman spectra. Also, all these measurements were done immediately after the sample cultivation. Raman microscope (Bio Particle Explorer, rap.ID Particle Systems GmbH) comprised of Nd:YAG laser (LCM-s-111-NNP25, Laser-export Company *Ltd.*) operating at 532 nm is utilized in the experiment. A 100x objective lens (MPLFLN-BD, Olympus) was used to focus the laser beam onto the nickel foil. It offers a lateral resolution of less than 1 μm with a maximum laser output of ~3 mW. The backscattered light from the sample was focused onto a single-stage monochromator (HE 532, Horiba). It is equipped with a grating of 920 lines/mm and coupled to a thermoelectrically cooled CCD detector (DV401A-BV, Andor). The resulting spectral resolution was around ~ 8 $cm^{-1}$. Finally, a total of 5420 spectra were acquired in these conditions, and the raw Raman spectra are presented in Supplementary Fig. S1.

*2.2 NRB simulation details*

The NRB is simulated as a linear function with different strengths. In this context, 'Strength' represents the intensity of the NRB relative to the resonant Raman signal, where its maximum value is set to 1. Therefore, the NRB strength is systematically varied relative to the Raman signal by considering the following values: 0.01, 0.1, 1, 10, 100, and 1000, with each step increasing the strength by an order of magnitude. Additionally, for each strength ($\mu$), the NRB is generated within a specific variance ($\sigma$) value, assuming a Gaussian distribution. For example, a typical schematic of NRBs simulated with $\mu = 100$ and $\sigma = 1$ is shown in Fig. 1(a). The dotted sky-blue line at 100 represents the NRB strength, and the dotted red lines approximately correspond to the $3\sigma$ variance of the Gaussian distribution. The NRB is generated as random lines with different slopes and intercept values within this variance region. The wavenumber scale is flipped after each iteration to obtain the negative slope for the NRB, ensuring that all possible variations within the selected variance are captured, as depicted in Fig. 1(a).

Furthermore, the variance values significantly influence the level of fluctuations present in the simulated NRB, thereby impacting the overall shape and intensity of the final CARS signal. Hence, the variance is varied in the range of 0.001 to 100 for each NRB strength. However, it's important to note that when the strength is less than or equal to the variance ($\mu \leq \sigma$), it may result in negative NRB values in the Gaussian distribution, which are physically meaningless in CARS spectroscopy. Consequently, such combinations (e.g., $\mu = 0.1$ and $\sigma = 0.1, 1, 10, 100$) are omitted from consideration, resulting in a total of 21 possible combinations, as depicted in Fig. 1(b) with an asterisk mark (*). Finally, the NRBs are generated for all these 21 combinations by following the same procedure and subsequently utilized for the CARS data simulation.

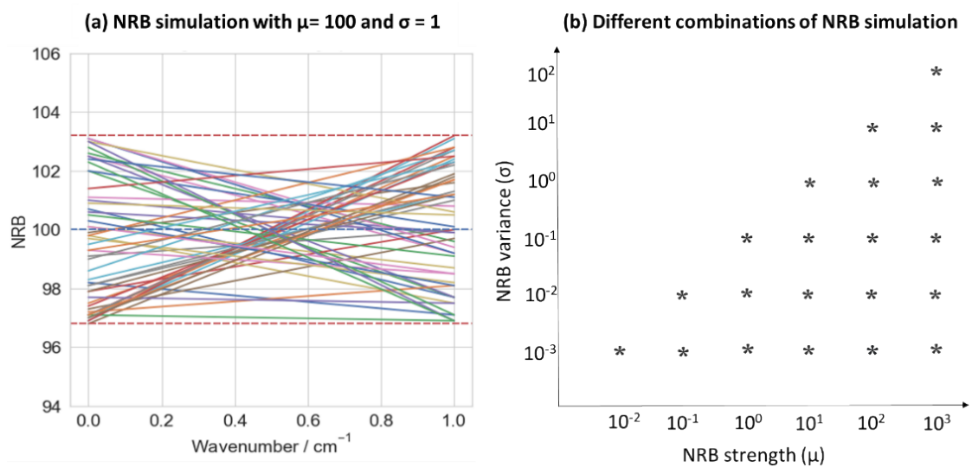

**Fig. 1.** a) Typical schematic of the NRBs simulation with strength of 100 and variance of 1. The dotted sky-blue line at 100 represents the NRB strength, and the dotted red lines correspond to the 3σ variance approximately. The NRB is generated within this range with different slopes and intercept values. b) Different combinations (*) of strengths and variances are considered for NRB simulation for the present study.

## 2.3 CARS data simulation details

The simulation of the CARS signal involves both resonant and non-resonant components, as described in equations 1 and 2. The NRB simulation process has already been detailed in the previous section. In this section, the resonant signal is simulated from the experimental Raman spectra acquired from six different bacteria across nine independent batches. These spectra exhibit considerable batch-to-batch variations commonly encountered in biomedical applications. Consequently, the simulated CARS spectra can better approximate the experimental CARS spectra, and the data can be considered semisynthetic. First, the real part of the resonant signal is extracted from the raw Raman data using the KK relation[30]. However, it exhibited an erroneous shape at the ends. Hence, the input Raman data is background corrected using the statistics-sensitive nonlinear iterative peak-clipping (SNIP) algorithm, followed by min-max normalization.

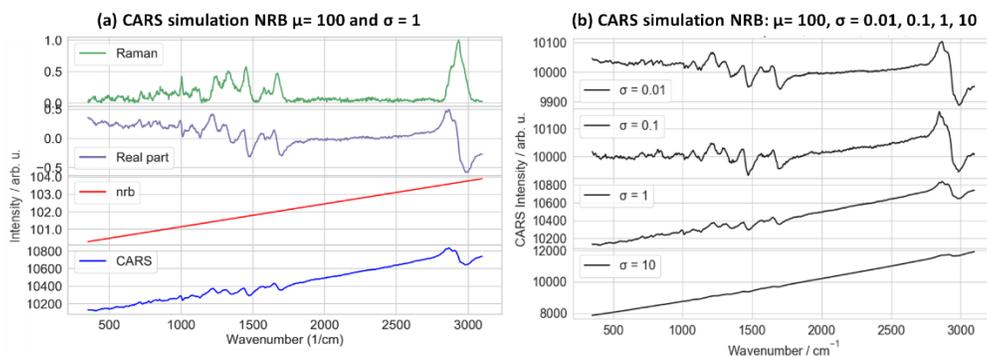

**Fig. 2.** a) Sequential steps involved in the CARS data simulation. Raman spectrum background is corrected and normalized (green). The corresponding real part was extracted using the KK

algorithm (purple). NRB spectrum simulated with a strength of 100 and variance of 1 (red). Final simulated CARS spectrum (blue). b) CARS data simulated with NRB strength 100 and variances of 0.01 (Top), 0.1, 1, and 10 (bottom).

The processed Raman spectrum and its corresponding real part are depicted in the first two rows of Fig. 2a, respectively. The real part now has proper lines shapes. Finally, CARS data is simulated using specific NRB strength and variance combinations. For instance, the CARS spectrum simulated with NRB $\mu = 100$ and $\sigma = 1$ is shown in Fig. 2(a) with a blue line. This procedure is repeated for all 21 possible combinations, as illustrated in Fig 1(b).

Further, the level of variance significantly influences the resulting CARS spectra for each NRB strength. For instance, Fig. 2(b) illustrates the CARS spectra simulated at NRB $\mu = 100$ with $\sigma = 0.01$ (top), 0.1, 1, and 100 (bottom). The positive slopes are considered solely for better visualization purposes. A systematic variation in the spectra is observed with increasing variance, as illustrated in the figure. Higher variances tend to hinder or even obscure spectral features, as evidenced by the last row of Fig. 2(b). In contrast, spectra simulated with lower variances exhibit only subtle differences, as depicted in the first two rows of Fig. 2(b). A similar trend is noted for CARS spectra simulated with NRB strengths of $\mu = 100$ and 1000 and variances of $\sigma = 0.001$, 0.01, and 0.1. These observations suggest that the NRB simulated at very low variance and strength approximates it as a constant value compared to the resonant signal as assumed in the literature[27].

## 3. Results and discussion

This section discusses the results obtained from the four different methods: MEM, KK, CNN, and LSTM. Initially, these methods are employed to retrieve the Raman signal from the simulated CARS spectra. Subsequently, the predictive performance is evaluated using Pearson correlation analysis, which compares the predicted output with the ground truth. Furthermore, PCA-LDA analysis is conducted to estimate the classification accuracy. In the text, the following terminology is utilized to represent the NRB strength and variance: "NRB of $10 \pm 1$" indicates simulation with a strength of 10 and variance of 1.

### 3.1 Extraction the imaginary part

Fig. 3(a-d) represents the imaginary part predicted by the MEM, KK, CNN, and LSTM models, respectively. The CARS spectrum simulated with the NRB of $100 \pm 1$ is used as input to the models. This spectrum is arbitrarily considered as an example from the entire data set only for visualizing the efficacy of the four methods. In each plot, the input CARS spectrum is presented at the top with a blue line. The predicted output and ground truth are shown in the middle, with red and green lines. It is worth noting that the different intensity scales are considered for ground truth and prediction, which helps readers to interpret the data more efficiently. Their absolute difference is estimated and presented with the black line at the bottom. This error plot serves as a visualization tool for evaluating the predictive performance of the four methods. All the methods except for LSTM have successfully retrieved the total spectral lines from the CARS spectra, as shown in Fig. 3(a-d). It is observed that MEM and KK predictions have been dominated by a significant background and retrieved spectral features are hindered by it. In the case of CNN, the background is minimal, albeit the output is noisy compared to others. The LSTM model consistently failed to extract the spectral features, resulting in a straight line with spurious lines at random places, as shown in Fig. 3(d). A similar behavior was demonstrated for LSTM when dealing with the real experimental CARS spectra[28]. This could be due to the simple architecture of the LSTM model.

Despite differences in the absolute intensity values, the predicted line shapes, including the background, are almost similar for MEM and KK methods. Also, the presence of a broad background contributes to a notable increase in error, as evidenced by the results shown in the last rows of Fig. 3 (a & b). Hence, the background from the extracted output is removed by applying the SNIP algorithm to minimize this error, and the results of the four models are presented in Fig. 3(e-h), respectively. After removing the background, the predicted output

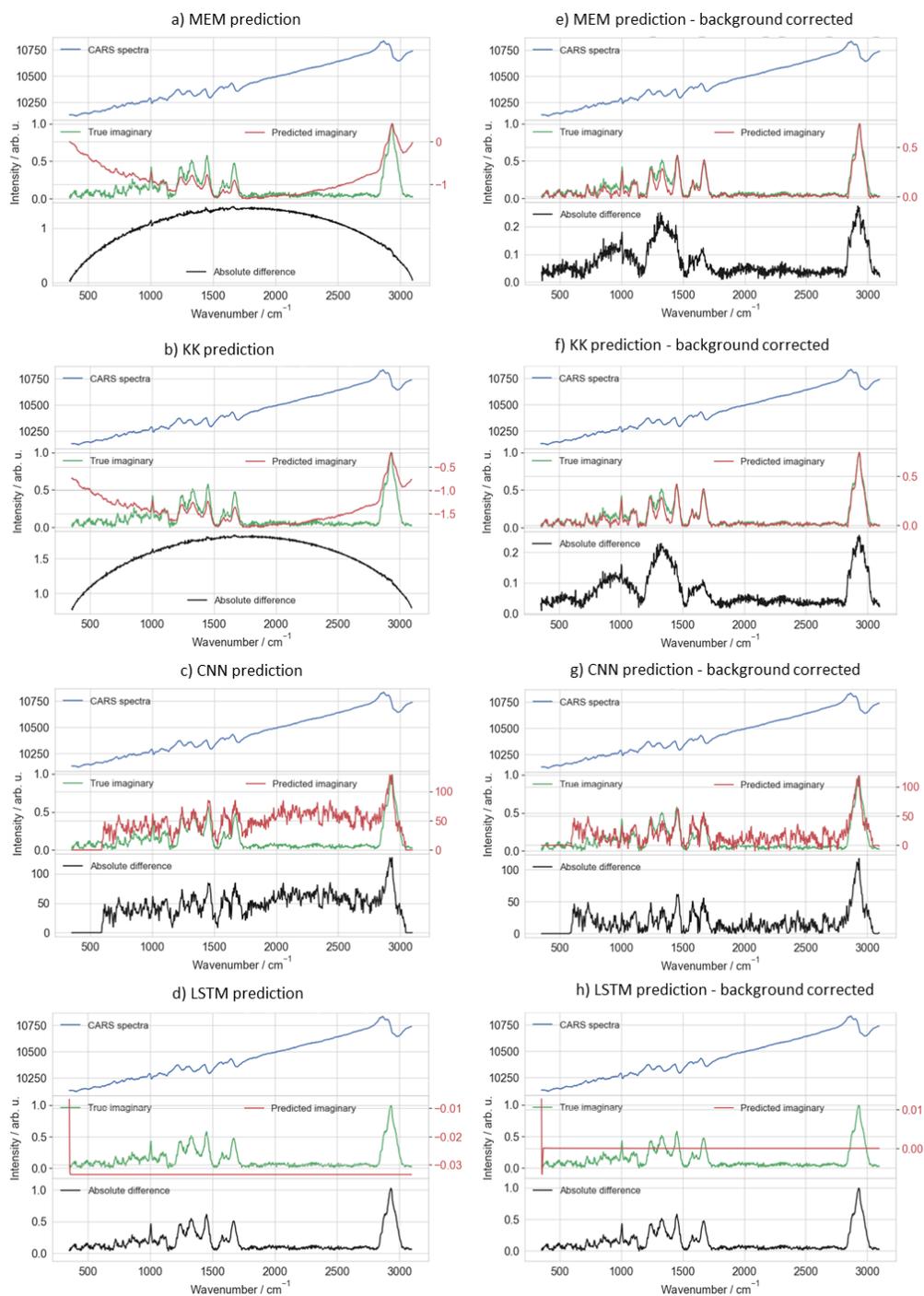

**Fig. 3**. a – d) Represents the predictions of the MEM, KK, CNN, and LSTM models, respectively. The input to models is the raw CARS data (blue line) simulated with NRB of 100

± 1. The middle row in each figure represents the predicted imaginary part (red line) and ground truth (green line). The last row visualizes the absolute difference between the predicted output and ground truth. The prediction was found to be superior for MEM and KK, followed by CNN and LSTM models. e-f): Represents background corrected imaginary part for the four models, respectively. The SNIP algorithm is used to remove the background.

closely resembles the ground truth with negligible variations in intensities. All the spectral lines, even with minute intensities, are correctly retrieved by the MEM and KK methods, as shown in Fig. 3(e & f). The background removal improved the CNN performance, but the intensities of the few retrieved lines deviated from the actual one. In addition, the noisy output limits its performance. This limited efficiency of the DL models could be due to the original models being trained with the 0-1 normalized CARS data as input[25], but raw CARS data is given as input here. Hence, the input raw CARS data is normalized between 0-1 to tackle the issue, and the retrieved imaginary parts are presented in supplementary Fig. S2(c). The normalization improved the efficiency of the CNN model, which removed the noisy pattern, and no background was observed. In the case of LSTM, some peaks in the fingerprint region are properly identified, as shown in supplementary Fig. S2(d). However, MEM and KK predictions still have the background, as shown in supplementary Fig. S2(a & b). Hence, the background is removed, and the results are presented in supplementary Fig. S2(e & f). Nevertheless, the extracted line intensities deviated from ground truth compared to raw data as input (see Fig. 3 e & f). It suggests that raw data as an input is best for MEM and KK, whereas normalized data gives optimum results for CNN and LSTM.

Further correlation analysis is performed in the next section to estimate the efficacy of the four models.

*3.2 Pearson correlation analysis*

Pearson correlation is a statistical technique that measures the strength of the linear relationship between two sets of continuous variables and provides a single numerical value, i.e., Pearson correlation coefficient (PCC). Here, it represents the similarity percentage between actual and predicted imaginary parts. The measured PCC values can vary between 100 and -100, representing positive and negative linear correlations, respectively. Overall, PCC value 100 represents the best prediction; the predicted imaginary part is identical to the ground truth, whereas 0 corresponds to no similarity. The PCC value is computed for each test spectrum for all four models, which gives a unique value for the predictions. Therefore, it can be considered as a performance metric for evaluating the predictive performance of all the models. First, the PCC values are computed for all the predictions of one dataset (ex: CARS spectra simulated with NRB of $100 \pm 1$) without any background correction, and their average is calculated. The same procedure is repeated for all the combinations, and results are presented as a heat map. For example, the results of the MEM predictions without background correction are visualized in Fig. 4(a). The x and y axes correspond to NRB strength and variance, respectively, and the heat scale (0-100) represents the average PCC value obtained for each combination. For example, the first box on the bottom left corresponds to the average PCC value estimated for NRB of $0.01 \pm 0.001$. A similar analysis is done for the remaining three models, KK, CNN, and LSTM, and the results are shown in Fig. 4(b-d), respectively.

It is noticed that the PCCs of MEM predictions are decreasing with increasing variance for each NRB strength, as shown in Fig. 4(a). It directly correlates that the predictive performance decreases with increasing variance. However, the change is drastic for the variance of 1, irrespective of the NRB strength, where PCC values are reduced to half of its previous variance. For example, the PCC value is 93.49 at NRB of $100 \pm 0.1$ but suddenly dropped to 42.14 for NRB of $100 \pm 1$. It suggests that NRB simulated with a variance of 1 significantly affects predictive performance. The efficiency further deteriorated with its increase, as shown in Fig. 4(a). In addition, a different scenario is observed for lower NRB strength (0.01) where the negative PCCs are achieved. It is merely due to the inverted shape after 2000 $cm^{-1}$ in the predicted imaginary part, as shown in Supplementary Fig. S3(a). The CH band in the 2800 – 3100 cm$^{-1}$ region mainly contributed to a negative correlation. A similar trend is observed for the KK predictions, and the measured PCCs differ by less than 10 percent compared to MEM, as shown in Fig. 4(b). However, a different scenario is observed for CNN, where the PCCs of lower NRB strength have shown a positive correlation with more than 80 % similarity. The extracted line shapes are presented in supplementary Fig. S3(b). It is evident from the figure that the CNN prediction works well with the low NRB strengths. However, the performance greatly decreased for higher NRB strengths, as shown in Fig. 4(c). Further, the LSTM model prediction was only good for the first 2 NRB strengths with more than 80 % positive correlation, whereas it is less than 10 % for the remaining.

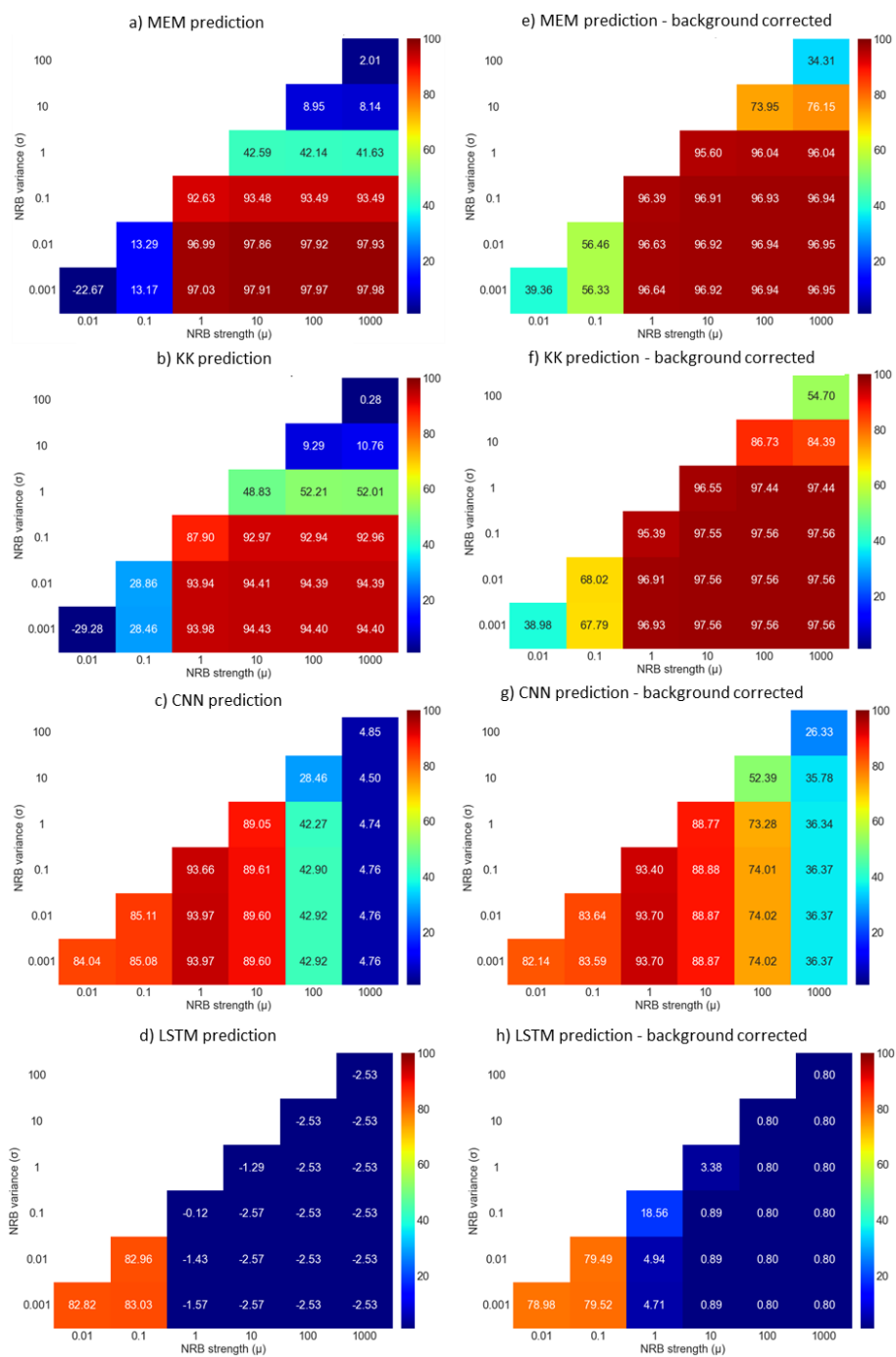

**Fig. 4.** a – d) Average of Pearson correlation coefficients (PCC) obtained from the MEM, KK, CNN, and LSTM predictions, respectively. Raw CARS data is used as input to the models. Higher PCC values were achieved for MEM and KK, followed by CNN and LSTM. Also, a negative correlation was observed for MEM and KK prediction for NRB 0.01 ± 0.001 due to the inverted shape after 2000 cm$^{-1}$ in the predicted imaginary part. e – h) PCC values obtained after removing the background from the predictions. The correlation increased after background removal but did not affect the LSTM predictions.

The PCCs are also estimated after removing the background from the predicted imaginary part, and the results are presented in Fig. 4(e-h). The correlation increased for all the models except for the LSTM. Also, it primarily impacted the predictions at higher NRB strengths. For example, the PCC is only 8.95 % for MEM prediction at NRB 100 ± 10 and increased to 73.95 % after background removal. A similar trend is noticed for the KK prediction, with a relatively higher correlation of more than 10 % compared to MEM. In the case of CNN, the correlation is increased around six times for the NRB strength of 1000. However, the final PCCs are not even half of the MEM and KK predictions. Also, the LSTM predictions are not affected by the background correction; hence, the PCCs are not changed, as shown in Fig. 4(h).

As mentioned in the previous section, the CNN and LSTM predictions improved after normalizing the input CARS data. Hence, the PCCs are also estimated for the predictions with normalized CARS data as an input to the models, and the results are presented in Supplementary Fig. S4. The correlation is significantly increased for CNN and LSTM at higher NRB strengths as compared to raw data as an input. It is primarily attributed to the reduced noise in the predicted imaginary part along with no background, as shown in supplementary Fig. S2(c & d). In the case of MEM and KK, the correlation decreased due to background and noise in the predictions. Nevertheless, the correlation is later improved after background correction, as shown in Supplementary Fig. S4(e & f). However, it is still less than the raw data as input to models, as shown in Fig. 4(e & f). These observations manifest that the raw CARS data as input would be best for MEM and KK. In contrast, normalized data demonstrates optimum results for CNN and LSTM. Also, the predictive efficiency significantly improves after background correction for MEM and KK, and it would not be required for the other two methods.

Even though the PCC value can be considered as the best metric for comparing the predicted output with the ground truth, it only measures the strength of linear associations between pairs of continuous variables. Hence, a linear classification model is utilized in the next section, and its performance is used as a performance metric.

*3.3 Principal-component analysis (PCA) - linear-discriminant analysis (LDA)*

Principal-component analysis (PCA) followed by linear-discriminant analysis (LDA) is applied as a classification model to separate six bacteria species. PCA identifies uncorrelated linear combinations of multiple variables to simplify and reveal patterns that are hidden in predicted output. PCA, followed by LDA, was applied to estimate the classification accuracy of the model predictions[33, 34] using different inputs. PCA, being an unsupervised technique, reduces high dimensional input data into a lower dimensional space by retaining most of the information. In contrast, LDA is a supervised method applied on top of these initial steps, which provides a classification model and is evaluated using classification accuracy. PCA primarily projects datasets in the direction representing the largest possible variance, called principal components (PCs). All the PCs are arranged such that maximum variance is retained by the first PC, i.e., PC1, and followed by subsequent PCs. First, the PCA is applied to the model predictions and retrieved the PCs. Later, the LDA model is built by considering the first 10 PCs as input. The leave-one-out cross-validation (LOO-CV) method is used for the validation. Here, one batch is utilized as a validation set out of nine independent batches, and the classification model is constructed based on the remaining batches. The same procedure is repeated for all batches as a validation set, and finally, average accuracy is estimated. This validation approach ensures that the independent datasets are being used for training and validation while evaluating the reliable and unbiased estimate of model performance. The entire analysis is carried out using in-house written functions in Python 3.9. Fig. 5(a-c) represents the PCA loadings plot obtained from the MEM, KK, and CNN predictions, respectively. The three rows in each figure correspond to the loadings of the first three PCs. As seen from the figure, a huge background contributes to the first PC for all the models, whereas little background is observed in the - remaining PCs. In addition, the loadings look noisy for the CNN model. It is worth noting that all the LSTM model predictions were found to be the same for NRB strength of more than 10. Thus, the PCA-LDA could not fit the data, and the loadings are not visualized in Fig. 5. The same analysis repeated after removing the background from the predictions and corresponding loadings are illustrated in Fig. 5(d-f). Now, the baseline of all the loadings almost looks flat, and spectral features buried in the background before its removal have appeared in the first PC. As aforementioned, LDA was applied to the first 10 PCs, and accuracies obtained for each-model are visualized in the heat map, as shown in Fig. 6. For each NRB strength, the accuracies slowly decreased with increasing variance, but the change is noticeable for the variance of more than 0.1. It is observed that the measured classification accuracies are approximately the same for the MEM and KK, as illustrated in Fig. 6(a & b). In the case of CNN, the accuracies are slightly lower than MEM and KK except for the NRB of 1000 ± 100.

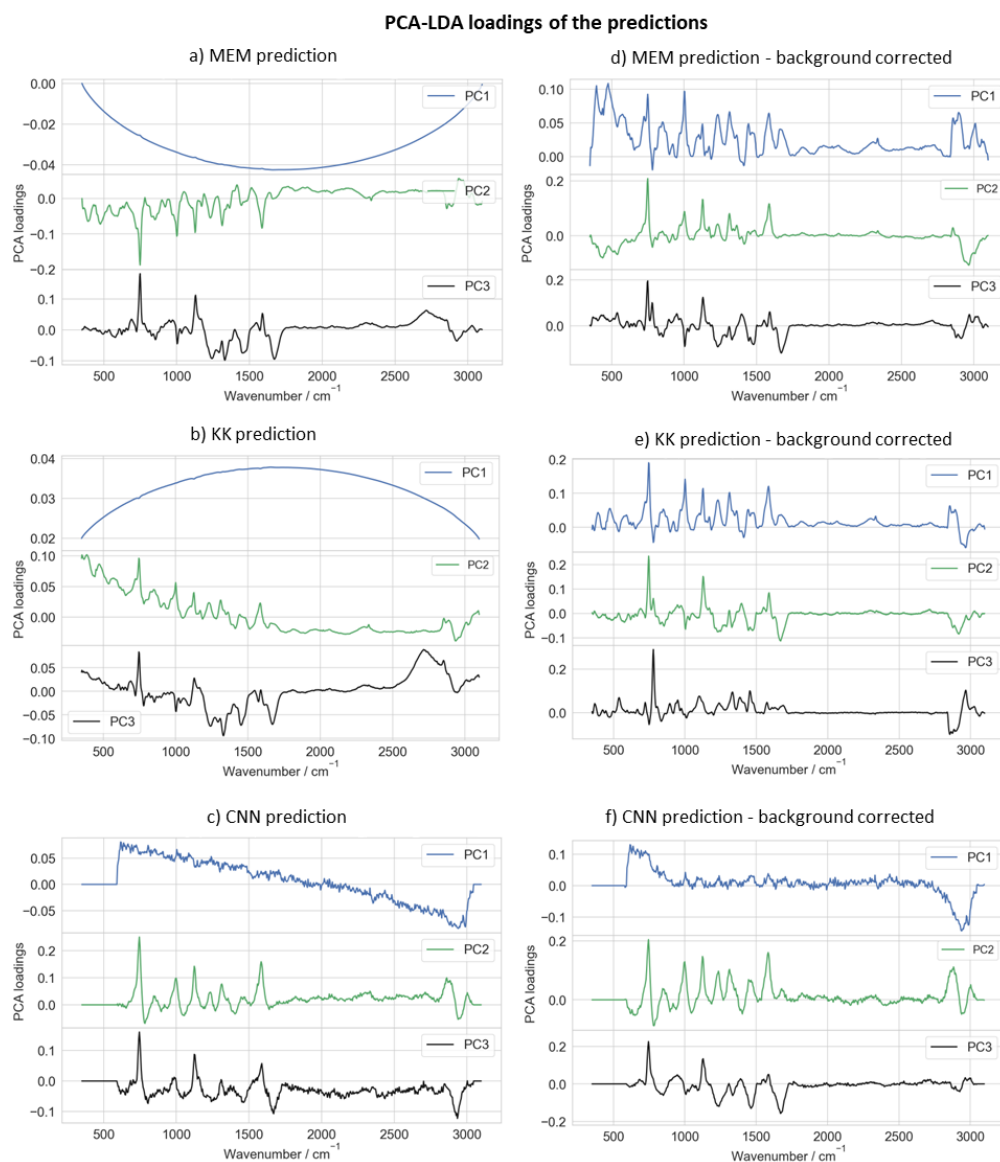

**Fig. 5.** a – c) Represent the PCA loadings plot obtained from the predictions of the four models, MEM, KK, and CNN, respectively. PC1, PC2, and PC3 correspond to the first three principal components (PCs). A large background is observed in the first PC compared to the remaining PCs for all the models. The LSTM model predictions were found to be the same for NRB strength of more than 10. Hence, the PCA-LDA could not fit the data, and the loadings are not visualized here. d – f) Represents the PCA loadings plot obtained after removing the background from the predictions of MEM, KK, and CNN models, respectively. The spectral features buried in the spectra have appeared in the first PCs after removing the background.

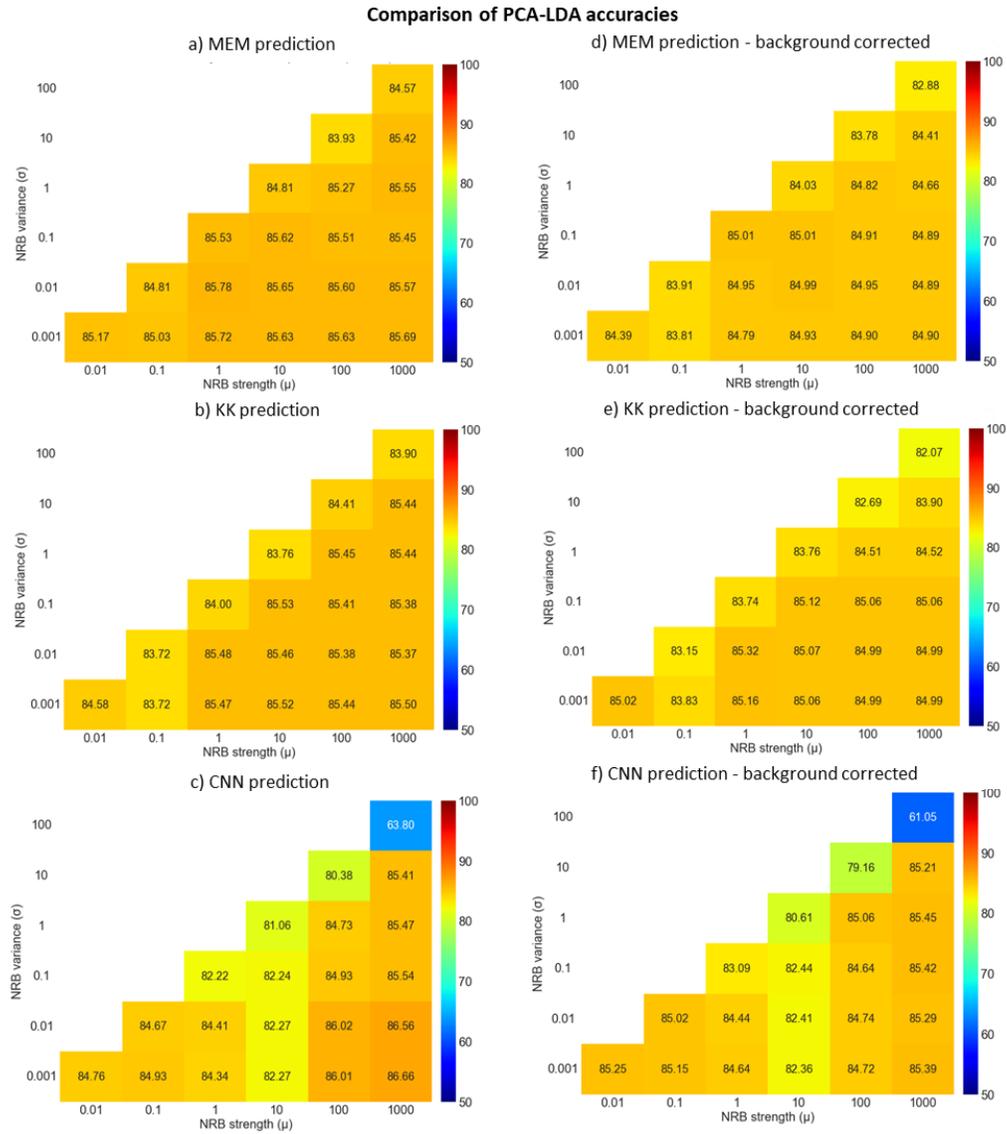

**Fig. 6.** (a–c) Represent the PCA-LDA classification accuracies obtained MEM, KK, and CNN predictions, respectively. Here, raw CARS data is used as an input to the models. The accuracies for a particular NRB strength decreased slowly with increasing variance. The accuracies of MEM and KK predictions are approximately the same. (e – f) Represents the classification accuracies obtained after removing the background from the predictions. Higher accuracies were achieved for MEM and KK, followed by CNN. Also, the classification accuracy is found to be the same even after background correction.

Also, no significant differences were observed after removing the background from the predictions, as shown in Fig. 6 (d & e). The change in accuracy is also found to be less than 1.5 % for all the combinations except for the NRB of 1000 ± 100. It could be due to the LDA model built based on the PCs not having any background, as shown in the coefficient plot in Supplementary Fig. S5. It suggests that the PCA-LDA model can classify samples even with a background whose classification accuracies are close to predictions after background correction. The same analysis is repeated for the predictions obtained for 0-1 normalized CARS data as an input to the models. The loadings plot is shown in Supplementary Fig. S6, and corresponding accuracies are shown in Supplementary Fig. S7. The accuracies decreased compared to raw data as input. The reduction is less than 4 % for all the NRB strengths up to 0.1 variance. However, the reduction is 15-20 % for the variances $\geq 1$.

All these results suggest that better prediction capability of MEM and KK models can be achieved with raw data as input. In contrast, normalized data would be best for obtaining optimum results with CNN and LSTM models. Also, PCA-LDA analysis revealed that it could be able to well classify the predictions even with background. The overall classification accuracies vary between 80 – 85 % for the MEM and KK methods and 50-85 % for the CNN and LSTM models. Especially, the performance of MEM and KK is found to be superior, particularly at higher NRB strength and variance. This capability plays a prominent role while working with real experimental CARS spectra for different applications. In future work, different types of NRBs will be explored. Also, the analysis would extend to other DL algorithms which have been explored recently.

## 4. Conclusions

In this work, we demonstrated the systematic study of NRB variation on CARS data analysis. The CARS data was generated with different NRB strengths by assuming it as a linear function. The NRB strengths were varied in the range of 0.01 - 1000 in the steps of increasing an order each time. Different variance levels were also considered for each strength. During the CARS data simulation, the real part of the resonant nonlinear susceptibly term was extracted from the experimental Raman spectra using the KK relation. Then, the imaginary part of the simulated CARS spectrum was extracted using four methods: MEM, KK, CNN, and LSTM. Finally, the model performances at different NRB strengths are evaluated based on correlation and PCA-LDA analysis.

First, the Pearson correlation analysis revealed that the prediction capability is approximately the same for the MEM and KK, followed by CNN and LSTM. The limited performance of LSTM could be due to the simple architecture of the model. Further, the Pearson correlation of the MEM and KK predictions was increased after removing the background from the extracted imaginary part. In contrast, normalizing the input data favored the CNN and LSTM model performance, whereas it negatively impacted the remaining two methods. Fig. 7(a) visualizes the final overview of the best methods suited for each NRB strength combination based on measured PCC values. In which, MEM prediction on raw CARS data is best for NRB strength $\geq 1$ and variance < 0.1, whereas KK predictions with background correction are optimum for higher NRB strengths. The CNN method is suited for very low NRB strengths.

PCA-LDA analysis demonstrated that higher accuracies are achieved for MEM and KK, followed by CNN and LSTM. It also revealed that the classification accuracies are not affected by the presence of the background in the predicted imaginary part. Fig. 7(b) illustrates the overview of the best methods for each NRB combination based on PCA-LDA accuracies. The MEM and KK models' prediction on raw CARS data is best for NRB strength $\geq 0.1$. The CNN predictions with background correction are optimum for NRB strengths < 1.

For the first time, this study systematically investigates how different NRB correction techniques perform if different NRBs (strength and variance) are found in the data. This should give a roadmap of which methods should be applied to the given experimental data at hand. As the simulation was only using linear NRBs, other functions need to be investigated in the future.

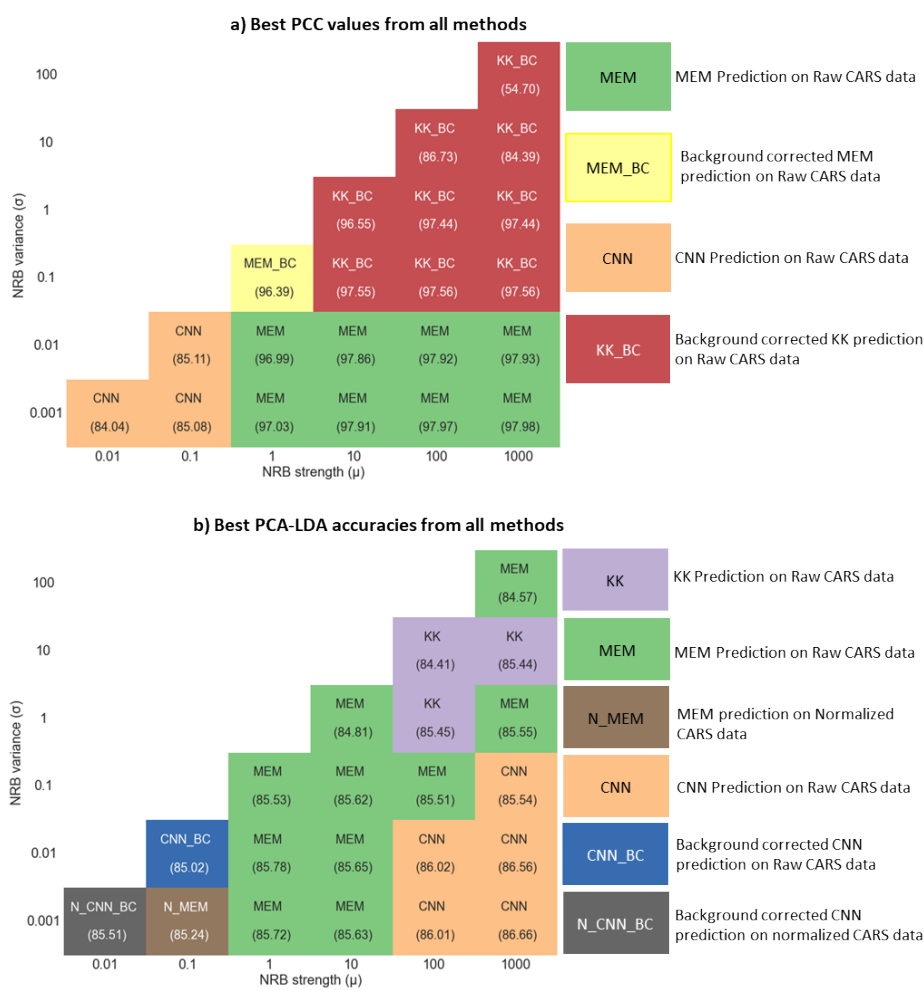

**Fig. 7.** (a) Represent the best PCC values obtained from all the methods for each NRB strength combination. It is noticed that MEM prediction on raw CARS data is best for NRB strength ≥ 1 and variance < 0.1. In the case of higher NRB strengths and variance. The KK model predictions with background correction are optimum. The CNN model is suited for very low NRB values. (b) Represents the best classification accuracies obtained from all the methods for each NRB strength combination. Higher accuracies were achieved for MEM and KK predictions on raw CARS data for higher NRB strengths and variance.

**Funding.** This work is supported by the EU funding program with grant number 101016923 for the project CRIMSON. This work is supported by the BMBF, funding program Photonics Research Germany (13N15466 (LPI-BT1), 13N15706 (LPI-BT2)) and is integrated into the Leibniz Center for Photonics in Infection Research (LPI). The LPI initiated by Leibniz-IPHT, Leibniz-HKI, Friedrich Schiller University Jena and Jena University Hospital is part of the BMBF national roadmap for research infrastructures.

**Disclosures.** The authors declare no conflicts of interest.

**Data availability.** Data underlying the results presented in this paper are not publicly available at this time but may be obtained from the authors upon reasonable request.

**Supplemental document.** See Supplement 1 for supporting content.